\documentclass[11pt,twocolumn,twoside,a4paper,amsmath,amssymb,aps,showkeys,showpacs]{revtex4}

\usepackage{amsfonts}
\usepackage{graphics} 
\usepackage{epsfig}
\usepackage{fancyheadings}

\begin{document}
\thispagestyle{myheadings}
\rhead[]{}
\lhead[]{}
\chead[B. Boimska, H. Bia{\l}kowska]{Some soft aspects of relativistic ion collisions}

\title{Some soft aspects of relativistic ion collisions} 

\author{Bo{\.z}ena Boimska}
\email{boimska@fuw.edu.pl}

\author{Helena Bia{\l}kowska}

\affiliation{%
The Andrzej So{\l}tan Institute for Nuclear Studies, Ho{\.z}a 69, PL-00-681, Warsaw, POLAND }%

\received{?}

\begin{abstract}
Concepts of wounded nucleon and quark participants have been used for years to parametrize and/or to explain many features of high energy nuclear collisions. Some results illustrating successes and failures of these two approaches are presented, including the latest developments.
In particular, results on identified particle production from nuclear collisions measured by the NA49 experiment at the CERN-SPS are shown.  The study has been done for both the nucleon and the constituent quark frameworks using the nuclear overlap model.
In addition, some preliminary observations concerning the behavior of $p_{T}$ spectra at forward rapidities, expressed in terms of the nuclear modification factor, for hadron-nucleus collisions at the SPS energy are also presented. These results are in relevance to RHIC results for deuteron-gold collisions often interpreted as a manifestation of saturation and/or color glass condensate.
\end{abstract}



\maketitle

\section{Introduction}
\label{introd}
Studying high energy hadron-hadron, hadron-nucleus and nucleus-nucleus reactions provides an opportunity to investigate strong interactions. Of fundamental interest are bulk observables such as multiplicity and particle densities (spectra), which are sensitive tools for probing the dynamics of strong interactions. They give information about the system formed in these collisions. Comparative studies of different colliding systems (h+h, h+A, A+A) enable distinguishing between various particle production scenarios.

Some aspects of these comparisons are not well understood. This paper covers two such subjects. In Section~\ref{wnqm}, results on the most basic observable, i.e., {\bf multiplicity of particles}, from h+h, h+A and A+A reactions at different energies, are shown. There are predictions of the Wounded Nucleon and Wounded Quark Models confronted with the yields measured experimentally.  The evolution of transverse momentum ($p_{T}$) spectra, expressed in terms of a so-called {\bf nuclear modification factor}, with rapidity $y$ (or Feynman $x_{F}$) and energy is discussed in Section~\ref{nmf}. These results concern hadron-nucleus collisions and are in relevance to saturation and/or Golor Glass Condensate effects. Finally, Section~\ref{sum} summarizes the presented issues. 

\section{Particle multiplicity and predictions of Wounded Nucleon and Wounded Quark Models}
\label{wnqm}
\begin{figure}
\includegraphics[scale=0.42]{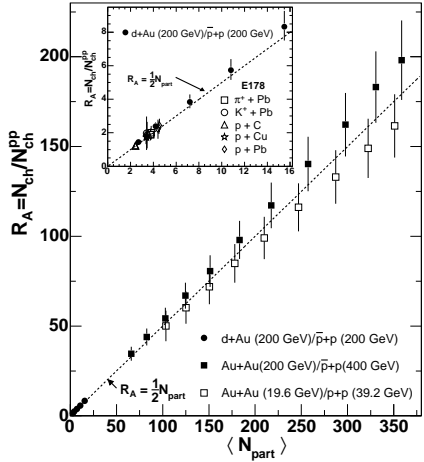}
\caption{\label{fig:hA_AA}Ratios of total particle multiplicity data for hadron-nucleus collisions (d+Au at $\sqrt{s_{NN}}$=200~GeV and $\pi$+A, K+A, and p+A at $\sqrt{s_{NN}}\approx$10-20~GeV) and nucleus-nucleus collisions  (Au+Au at $\sqrt{s_{NN}}$=19.6 and 200~GeV)  over the multiplicity in proton(antiproton)-proton interactions plotted  versus the number of participating nucleons. The denominator for h+A interactions is p(\=p)+p data at the same $\sqrt{s_{NN}}$. For Au+Au interactions, the denominator is  p(\=p)+p data at twice the center-of-mass energy.~\cite{bac05}}
\end{figure}

More than thirty years ago a concept of {\em wounded}\, nucleons was introduced~\cite{bia76}. A wounded nucleon (often called also a {\em participant}\,) is a nucleon that underwent at least one inelastic collision. The Wounded Nucleon Model (WNM) proposes  that  particle production in a nuclear collision is a superposition of independent contributions from the wounded nucleons in the projectile and in the target. Thus, the ratio of the number of charged particles produced in  h+A collisions ($N_{ch}^{hA}$) to that in p+p collisions ($N_{ch}^{pp}$) is:
\begin{equation}
R=\frac{N_{ch}^{hA}}{N_{ch}^{pp}}=\frac{N_{part}^{hA}}{2},
\end{equation}
where: $N_{part}^{hA}$ is the number of wounded nucleons (participants) in h+A , and ``\,2\,'' means that there are two participants in a proton-proton collision. A remarkable success of WNM is seen in  the inset of Figure~\ref{fig:hA_AA}, where  the total charged particle multiplicity from hadron-nucleus collisions is divided by the p+p multiplicity at the same collision energy. The ratio remains directly proportional to half the number of nucleon participants in h+A, as all points fall on the indicated line. 

A similar analysis of nucleus-nucleus data is shown in the main part of Figure~\ref{fig:hA_AA}. Again, the points fall along the line, exhibiting scaling of the total multiplicity with the number of participant pairs. In this case, however, WNM does not work as the reference p+p data are taken at twice the center-of-mass energy (to account for ``leading particle'' effect).
\begin{figure}
\includegraphics[scale=1.1]{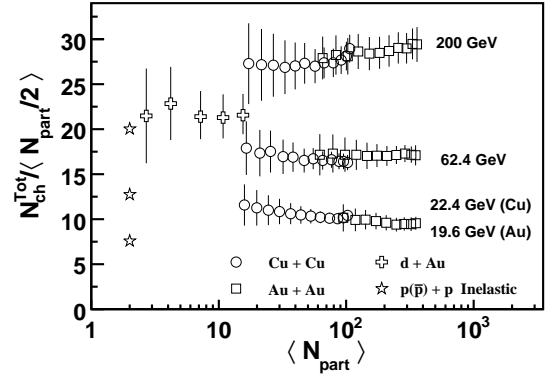}
\vspace*{-0.2cm}
\caption{\label{fig:tot_AA}Total charged particle multiplicity per participant pair as a function of number of participants. Data are shown for A+A  and p(\=p)+p collisions at $\sqrt{s_{NN}}$ of $\sim$20, 62.4 and 200~GeV, as well as d+Au at 200~GeV.~\cite{alv09}}
\end{figure}
Also Figure~\ref{fig:tot_AA} supports this statement. The total charged particle multiplicity per participant pair, for A+A, clearly exceeds the multiplicity from p+p collisions at the same energy. Still, the proportionality of these multiplicities to the number of nucleon participants holds, as a flat dependence on $N_{part}$ is seen.

\begin{figure}
\includegraphics[scale=0.6]{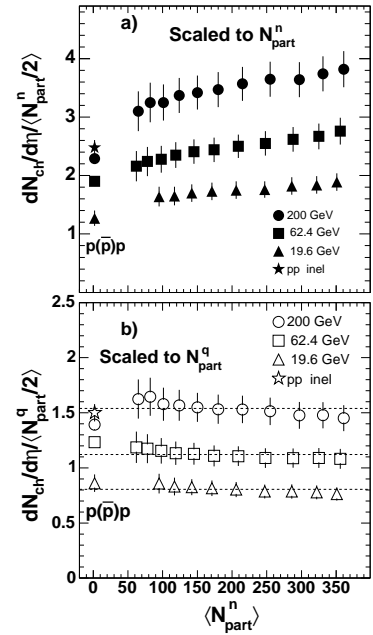}
\vspace*{-0.2cm}
\caption{\label{fig:quark_AA}Particle density at mid-rapidity, $dN_{ch}/d\eta$, divided by a)~the number of nucleon participant pairs, b)~the number of constituent-quark participant pairs for Au+Au and p(\={p})+p collisions at three energies.~\cite{nou07}}
\end{figure}
Charged particle densities ($dN_{ch}/d\eta$) per nucleon participant pair at mid-rapidity in A+A collisions are substantially higher than those in p+p collisions at the same energy, as shown in Figure~\ref{fig:quark_AA}a, pointing to the failure of WNM. In addition, there is the dependence on collision centrality  observed, reflecting the violation of $N_{part}$-scaling. One may ask here how the results change if quark participants are used in the normalization, instead of nucleon participants.

\begin{figure*}
\includegraphics[scale=0.25]{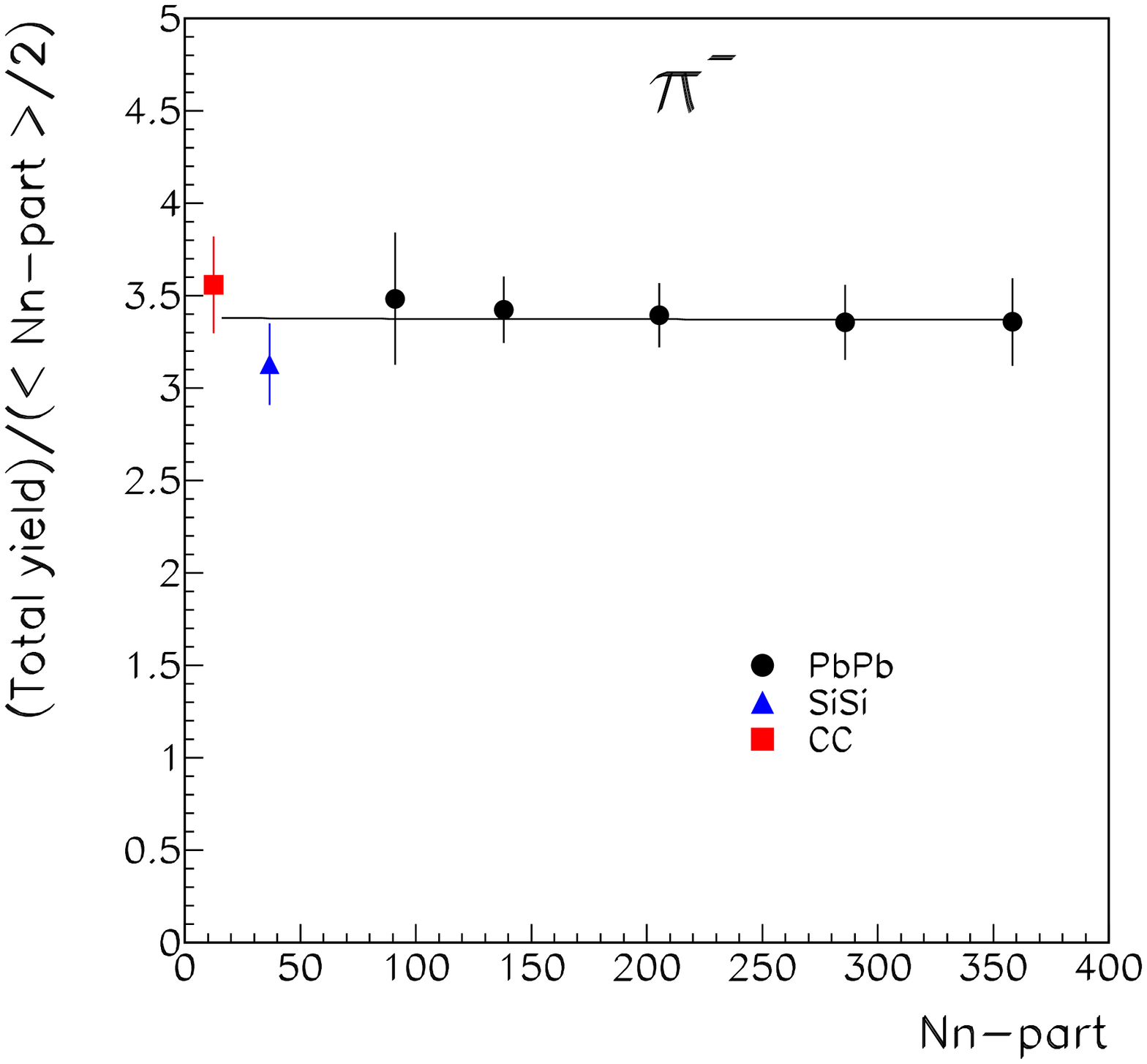}
\includegraphics[scale=0.25]{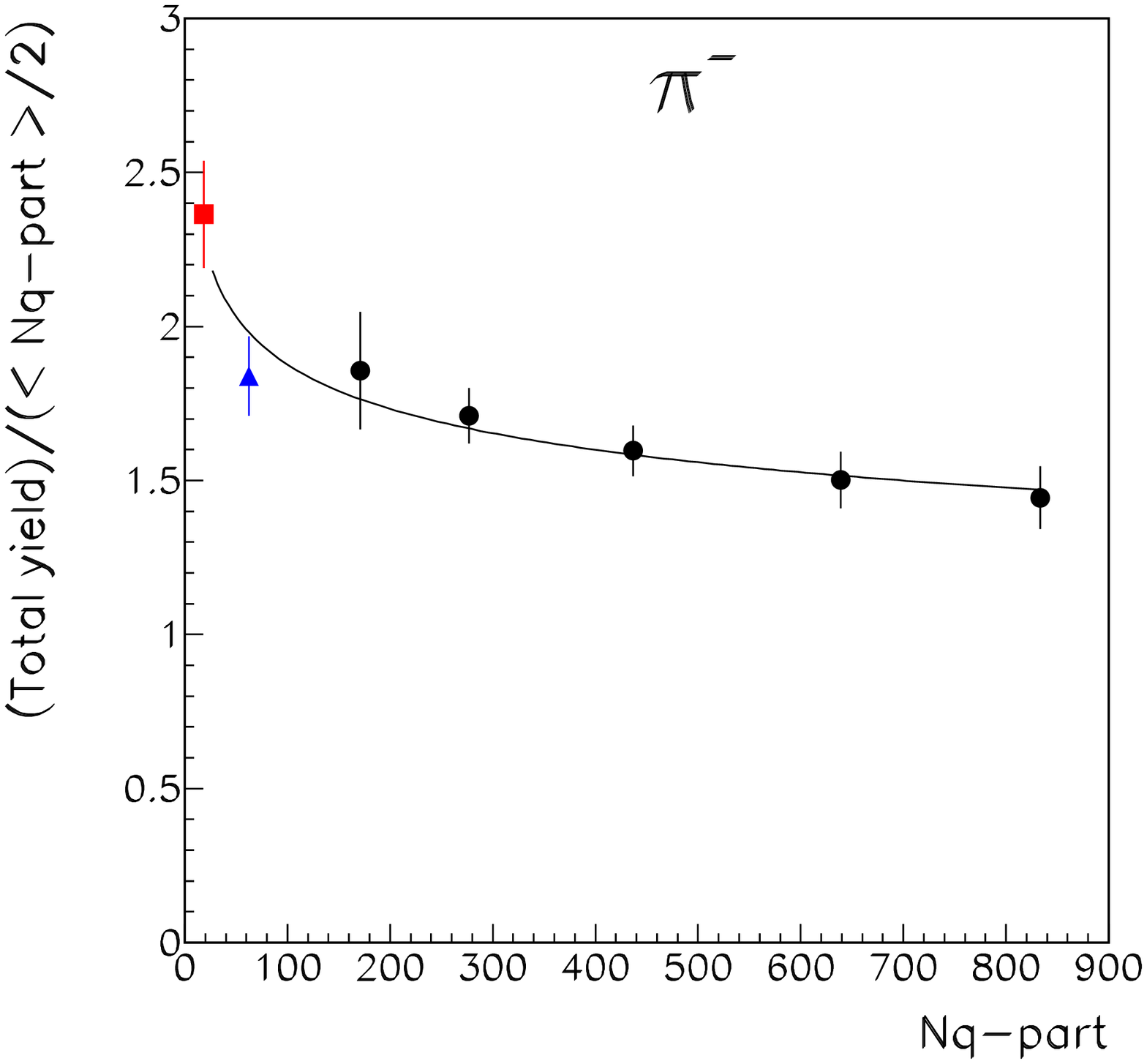}\\
\vspace*{-0.8cm}
\includegraphics[scale=0.25]{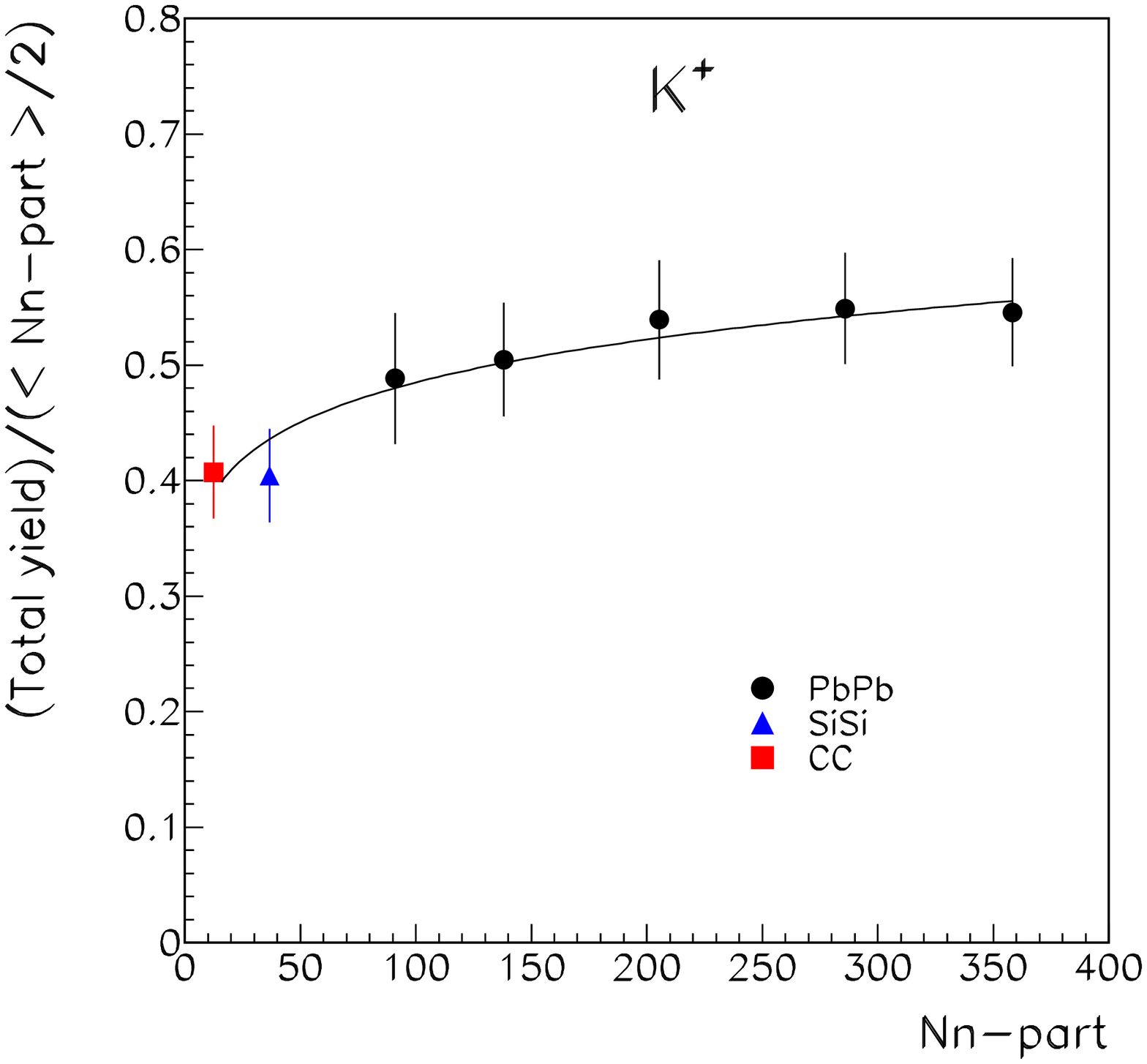}
\includegraphics[scale=0.25]{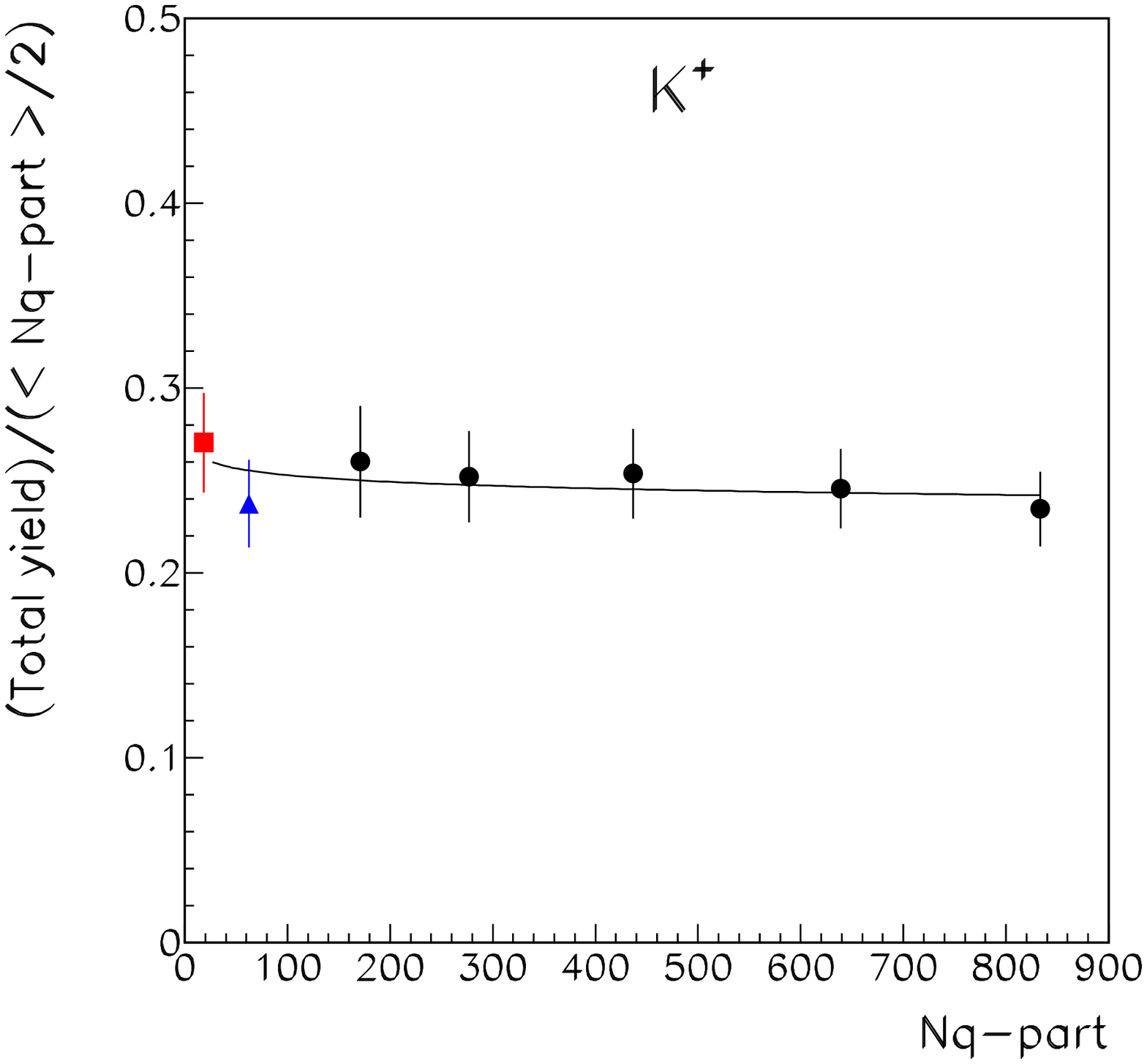}
\vspace*{-0.2cm}
\caption{\label{fig:id_prod_sps} Total yields of $\pi^{-}$ (upper panels) and K$^{+}$ (lower panels) for participant nucleon and quark scalings (left and right panels, respectively). Results are for C+C, Si+Si and Pb+Pb collisions at $\sqrt{s_{NN}}$=17.2~GeV.}
\end{figure*}

The concept of wounded quarks has been known for many years~\cite{bia77, ani78}. Results presented below are from the approach where both nuclei and single nucleons are considered as a superposition of constituent quarks; there are three such quarks per nucleon. The number of participating objects is estimated using the nuclear overlap model~\cite{esk89}. The nuclear density profile is assumed to have a Woods-Saxon form. In order to calculate the number of participating quarks ($N_{part}^{q}$) the density was increased three times and the nucleon-nucleon cross section was changed to the quark-quark one. When the A+A results for mid-rapidity particle density are normalized to the number of quark participant pairs (Figure~\ref{fig:quark_AA}b) the centrality dependence flattens out. This arises from the relative increase in the number of interacting quarks in more central collisions. The approximate scaling with $N_{part}^{q}$ is observed, which means that the Wounded Quark Model (WQM) works quite well here.

The same approach has also been applied to identified particle production at SPS and RHIC energies~\cite{bha05}. However, the treatment of the SPS data (NA49~\cite{bach99,ant04}) was inconsistent; the authors mixed the integrated yields with mid-rapidity yields for different particles. Therefore, a new analysis of the NA49 data has been performed~\cite{boi09}. Recent data on pion and kaon production in Pb+Pb collisions at $\sqrt{s_{NN}}$=17.2~GeV~\cite{din09} were studied. In addition, C+C and Si+Si collisions~\cite{alt05} were included in the study. As an example, the total $\pi^{-}$ yield was found to scale with the number of participating nucleons - $N_{part}^{n}$, and K$^{+}$ with quarks - $N_{part}^{q}$ (see Figure~\ref{fig:id_prod_sps}). Similar findings have been obtained for the NA35 data where production rates of h$^{-}$ appeared to be proportional to the number of participating  nucleons, and K$^{0}_{S}$ - to the number of wounded quarks~\cite{kad95}.

\vspace*{-0.2cm}
\section{Nuclear Modification Factor in forward region}
\label{nmf}
\begin{figure*}
\includegraphics[height=4.0cm,width=14.0cm]{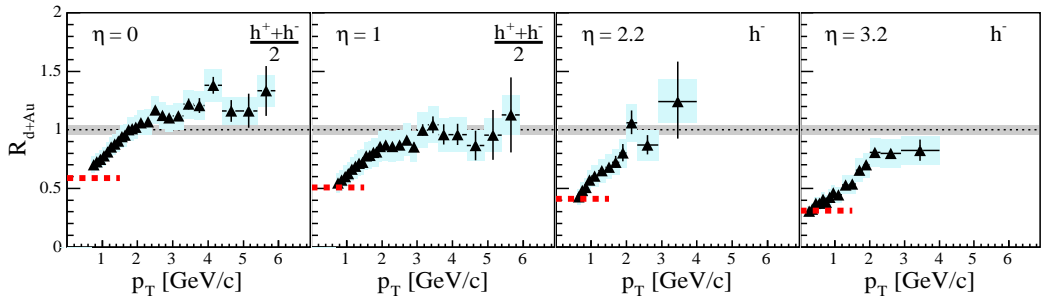}
\vspace*{-0.2cm}
\caption{\label{fig:rhic_dAu}Nuclear modification factor $R_{dAu}$ for charged hadrons at pseudorapidities $\eta$=0, 1, 2.2, 3.2.~\cite{ars04}}
\end{figure*}

Transverse momentum ($p_{T}$) spectra of particles produced in nuclear collisions are subject to various initial and final state effects. Initial state effects may enhance the yield - like the Cronin effect which strengthens it at intermediate-$p_{T}$ in p+A and A+A collisions as compared to p+p interactions, or suppress the yield - like nuclear shadowing and gluon saturation. Parton energy loss (final state effect), due to gluon {\em bremsstrahlung}\, during the passage through a very dense and colored medium created in ultra-relativistic A+A collisions, leads to the suppression of hadron yields at high-$p_{T}$ (jet-quenching). The degree of suppression/enhancement is quantified by means of the Nuclear Modification Factor (NMF). The initial state effects can be separated from the final state ones by studying  hadron-nucleus interactions, for which the latter effects are not present, and comparing to high energy nucleus-nucleus collisions.  
NMF for h+A collisions ($R_{hA}$) is defined as a ratio of the number of particles produced in a h+A collision over the number of particles produced in a p+p collision scaled by the number of binary collisions $N_{coll}$:
\begin{equation}
R_{hA}(p_{T},y)=\frac{1}{\left<N_{coll}\right>}\frac{d^2N^{hA}/dp_Tdy}{d^2N^{pp}/dp_Tdy},
\end{equation}
$N_{coll}$ - is the number of collisions the incident hadron would experience if it were not altered at all while passing through the nucleus A.

Quite recent measurements of hadrons produced in d+Au collisions at RHIC, $\sqrt{s_{NN}}$=200~GeV, showed  decreasing values of NMF with increasing (pseudo)rapidity of the particle~\cite{ars04,ada06} (see Figure~\ref{fig:rhic_dAu}). The observed suppression can be qualitatively described within the Color Glass Condensate (CGC) model~\cite{kha03a}, which assumes gluon saturation for the kinematic domain reached in hadron-nucleus collisions at RHIC. The saturation effects are expected to be more pronounced at large Feynman $x_{F}$, or large (pseudo)rapidity, where small-$x$ partons in the nucleus can be probed. The effects should also increase with the thickness of nuclear material traversed by the incoming probe, and indeed a greater suppression for more central d+Au collisions was observed~\cite{ars04}.

\begin{figure}
\includegraphics[height=6.0cm,width=7.0cm]{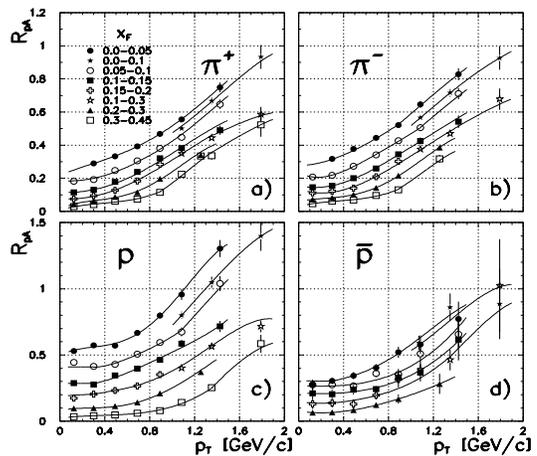}
\caption{\label{fig:sps_pA}Nuclear modification factor $R_{pA}$, in different $x_{F}$ intervals, for a)~$\pi^{+}$, b)~$\pi^{-}$, c)~$p$ and  d)~$\bar{p}$ produced in central p+Pb collisions at $\sqrt{s_{NN}}$=17.2~GeV.~\cite{boi04}}
\end{figure}

A similar suppression like at RHIC was measured also in p+Pb collisions at SPS, as shown in Figure~\ref{fig:sps_pA}. 
The measurements of $R_{pPb}$ by NA49~\cite{boi04} at an order of magnitude lower energy, where shadowing and gluon saturation are expected to be small, show qualitatively the same rapidity (or $x_F$) dependence. NMF decreases with  $x_F$ for pions as well as for p and \=p.

\begin{figure}
\includegraphics[height=4.5cm,width=5.0cm]{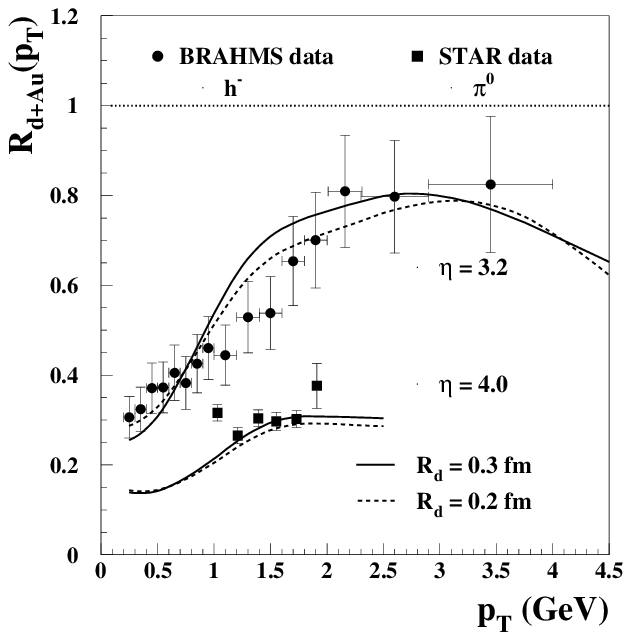}
\includegraphics[height=4.8cm,width=5.2cm]{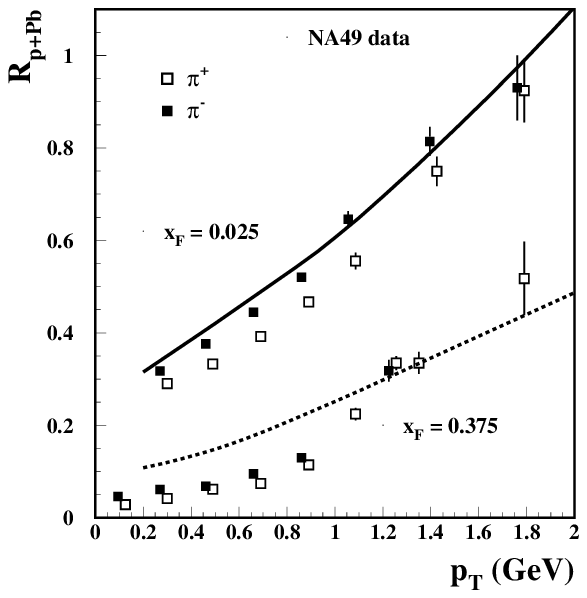}
\vspace*{-0.2cm}
\caption{\label{fig:rhic_sps_n} Model calculations and experimental data on NMF. Upper panel: $R_{dAu}$ for negative hadrons and neutral pions at pseudorapidity $\eta$=3.2 and 4.0~\cite{nem08}. Lower panel: $R_{pPb}$ for $\pi^{\pm}$ at $x_{F}$=0.025 and 0.375~\cite{nem09}.}
\end{figure}

The energy dependence of the suppression is related to underlying space-time dynamics of the collision and is therefore a crucial test for theoretical models.
Two approaches to study the energy dependence of the suppression (taking into account also the SPS result) were proposed~\cite{tyw07,nem08,nem09}. Both use the Glauber-Gribov model, but differ in technicalities. The studies were done in terms of ``standard'' effects, such as energy-momentum conservation, shadowing and Cronin effect. Effects related to Color Glass Condensate were not included in the models. The study within the first approach~\cite{tyw07} showed a good agreement with SPS and RHIC data on pion production. In addition, data at both energies seem to overlap (for $x_{F}$ dependence). This indicates that common mechanisms determine the suppression in the forward region both at SPS and RHIC. Figure~\ref{fig:rhic_sps_n} shows predictions of the second model~\cite{nem08,nem09}. The RHIC results are nicely described by the model (upper panel). This means, most likely, that a region where CGC-based models can fully be employed has not been reached at RHIC. The description of the SPS data is worse (lower panel).
\vspace*{-0.5cm}
\section{Summary}
\label{sum}
\noindent
-~~Some characteristics of particle production in high energy nuclear collisions scale with $N_{part}^{n}$, other with $N_{part}^{q}$. E.g., for A+A at SPS, the $\pi$ yield scales with  $N_{part}^{n}$ and K yield with  $N_{part}^{q}$.  
-~~Suppression in the forward region in h+A collisions is present not only at RHIC but also at SPS.
Understanding of the effect and its energy dependence needs more theoretical study.
\vspace*{-0.2cm}
\begin{acknowledgments}
B.\,B. would like to thank the organizers of ISMD 2009 for their hospitality during the conference.
\end{acknowledgments}

\end{document}